\documentclass{article}
\usepackage{amssymb}
\usepackage{algorithm}
\usepackage{algorithmic}
\usepackage{comment}
\usepackage{url}
\usepackage{spconf,amsmath,graphicx}

\title{Salient Building Outline Enhancement and Extraction using Iterative L0 Smoothing and Line Enhancing}
\name{Cho-Ying Wu and Ulrich Neumann}
\address{University of Southern California \\ Department of Computer Science}
\begin{document}
%
\maketitle
\begin{abstract}
In this paper, our goal is salient building outline enhancement and extraction from images taken from consumer cameras using $L_0$ smoothing. We address weak outlines and oversmoothing problem. Weak outlines are often undetected by edge extractors or easily smoothed out. We propose an iterative method, including the smoothing cell and sharpening cell. In the smoothing cell, we iteratively enlarge the smoothing level of the $L_0$ smoothing. In the sharpening cell, we use Hough Transform to extract lines, based on the assumption that salient outlines for buildings are usually straight, and enhance those extracted lines. Our goal is to enhance line structures and do the $L_0$ smoothing simultaneously. Also, we propose to create building masks from semantic segmentation using an encoder-decoder network. The masks filter out irrelevant edges. We also provide an evaluation dataset on this task.
\end{abstract}
\begin{keywords}
$L_0$ smoothing, buildings, enhancement, outline extraction
\end{keywords}
\section{Introduction}
\label{sec:intro}

In this paper, we focus on the salient building outline enhancement and extraction using image filtering. In the literature, building outline extraction is usually processed from satellite images or LiDAR data \cite{ortner2007building} \cite{gilani2016automatic}. However, extracted outlines are usually only outlines of rooftops. In this paper, we focus on the outline extraction from building images taken by a consumer camera. This new task is essential to image processing and has a lot of applications on architectural development, such as building exterior designs, building type classifications, architectural planning for areas, or 3D building reconstruction. Usually, building images from a consumer camera could contain a lot of irrelevant objects, such as trees, grasses, sky, and some small ornament objects. Also, materials of buildings contain structures. Thus, it is hard to extract building outlines from a single building image.

\begin{figure}[!htb]
    \centerline{\includegraphics[width=6.0cm]{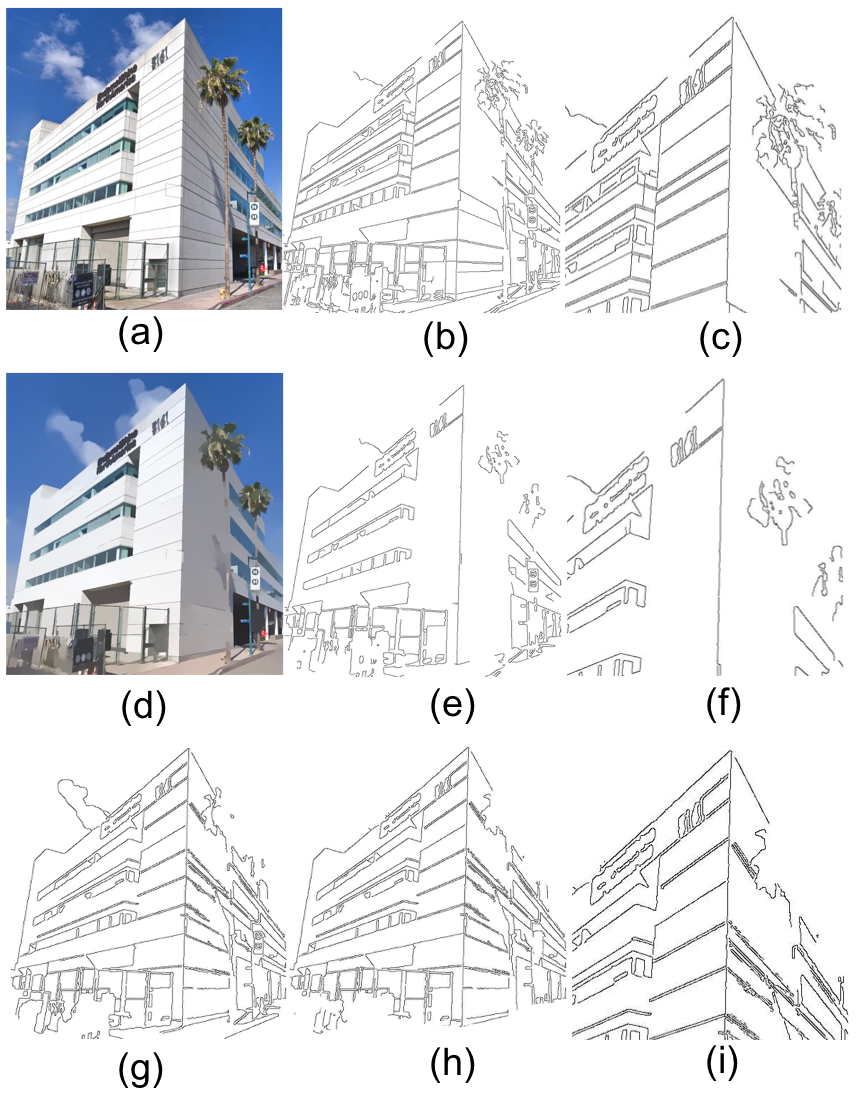}}
    \caption{The example of our method. All the edge maps are extracted by the Canny edge extractor with low and high threshold = $0.05$ and $0.3$ respectively. (a) The original example image. (b) The edge map of (a). (c) zoom-in map of (b). (d) The oversmoothing image of (a) using high $L_0$ smoothing level. (e) The edge map of (d). (f) zoom-in edge map of (e). (g) The recovered edge map using the proposed iterative smoothing line enhancing (ISLE) framework. (h) Apply (g) with the proposed building mask. (i) zoom-in edge map of (h). } 
    \label{fig1}
    \vspace{-0.3cm}
\end{figure}

Conventional edge extraction methods, such as Sobel, Prewitt, or Canny edge extractors \cite{gonzalez2002digital} could extract edges. However, the extracted edges are messy, including lots of non-building part edges, or excessive details of a building that could not represent building structures, like material textures. Some of the recent works use learning-based methods to predict the object salient outline. Oriented Edge Forest (OEF) \cite{hallman2015oriented} trains a random forest to predict outline structures. Bayesian Salient Edge \cite{mukherjee2017salprop} based on the OEF and further use Bayesian inference framework to refine the result. Richer convolutional features (RCF) \cite{liu2019richer} uses multi-stage networks to predict the salient outline. 

In this paper, we revisit edge-preserving filtering techniques. For example, bilateral filters \cite{tomasi1998bilateral}, total variance $L_1$ (TV-$L_1$) denoising filters \cite{beck2009fast}, and $L_0$ smoothing \cite{xu2011image} \cite{shen2012edge} are some well-known edge-preserving filters. In this paper we focus on the surprising smoothing ability of the $L_0$ smoothing. It could smooth out high frequency area, while preserving the low frequency area in an image. The $L_0$ smoothing technique minimizes the $L_0$ norm regularized squared difference of image recovery problem with a weight controlling the level of smoothing. The formulation is as follows.
\begin{equation}
\begin{aligned}
\label{L0}
\min_S \Big\{\sum_p(S_p-I_p)^2+\lambda\cdot C(S)\Big\},
\end{aligned}
\end{equation}
and $C(S)$ is
\begin{equation}
\begin{aligned}
\label{C}
C(S) = \#\{ p \; |\;|\partial_xS_p|+|\partial_yS_p| \ne 0\},
\end{aligned}
\end{equation}
where $I$ is the input image and $S$ is the optimized smoothed image. Subindex $p$ stands for the pixel $p$ in the image. $\partial_x$ is for the horizontal gradient operator, $\partial_y$ is for vertical gradient operator, and $\lambda$ is the weight.

If $\lambda$ is larger, the smoothing level is higher, leading to higher level abstraction. Thus, if we want to extract building outlines and smooth out internal area, higher smoothing level should be adopted. However, if the background or other irrelevant objects have similar pixel values with buildings, these weak boundaries could be smoothed out by using higher smoothing level. Thus, directly applying higher level smoothing could not retain salient outlines of building images.

We propose an iterative smoothing and line enhancing (ISLE) framework to simultaneously attain higher level smoothing and enhance salient outlines. Also, we propose to obtain building masks from an encoder-decoder structure network trained on ADE20K dataset \cite{zhou2017scene}. This building mask recognizes building area of an input image and help filter out non-building part edges. The example is in Fig. 1. (In this paper, thresholds of the Canny edge extractor is in $[0,1]$)

\section{Proposed Method}
\label{sec:format}

We address the weak edges and oversmoothing problem of the $L_0$ norm smoothing, as in Fig. 1. Our framework is an iterative method. Given an input image $I$, if the illumination is low, the edges are too weak to extract. Therefore, we first define 3 modes based on the histogram of the intensity map of the input image $I$. We set 2 thresholds $T_{high}$ and $T_{low}$. If the median of the intensity histogram, $med_I$, is larger than $T_{high}$, the illumination condition is good. We set the maximum iteration and threshold of Canny edge extractor to a higher value. If $med_I$ is between $T_{high}$ and $T_{low}$, then we turn on the medium mode to set both the maximum iteration and threshold of Canny edge extractor to a medium value. If $med_I$ is lower than $T_{low}$, then we turn on the low mode to set both the maximum iteration and threshold of Canny edge extractor to a low value.

\begin{figure}[!tb]
  \centerline{\includegraphics[width=8.0cm]{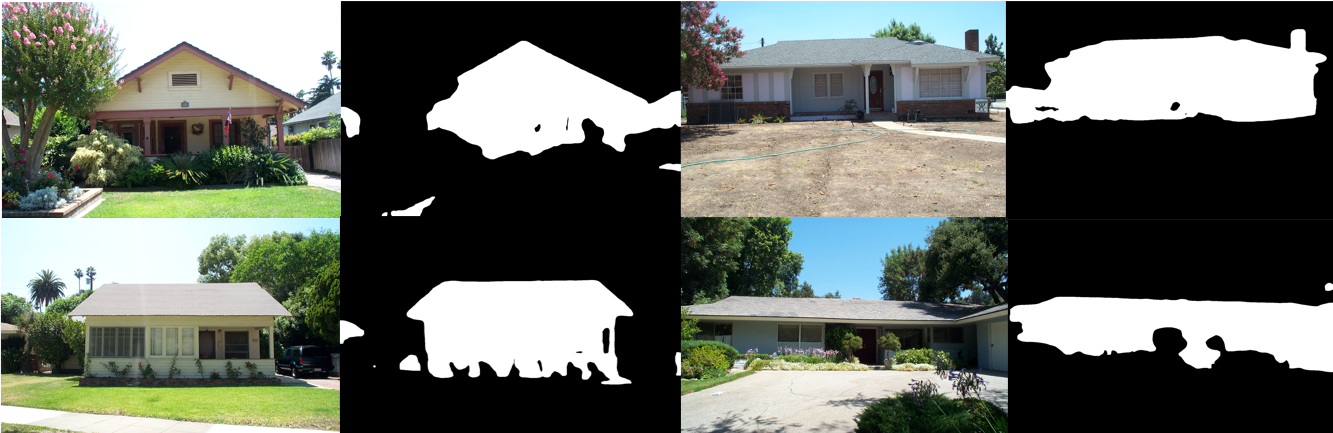}}
  \caption{Examples of the building mask obtained from our framework}
  \vspace{-0.3cm}
\end{figure}

Then we define two stages. The first is the smoothing cell. It contains $L_0$ smoothing to smooth the image. The smoothing level grows with iteration. That is, for the $k^{th}$ iteration, $\lambda_k < \lambda_{k+1}$.

The other is the sharpening cell. After the image has been smoothed, we use Canny edge extractor to extract the edges of the smoothed image. Since salient outlines of buildings are usually straight, if we can extract lines from the smoothed image and sharpen these line edges. These edges could be enhanced and possibly be robust to the smoothing in the later iterations. Therefore, we first use the Hough Transform to extract lines and mark these line positions. Then we do the line sharpening using a kernel as 
\begin{equation}
\begin{aligned}
\label{kernel}
ker = 
\begin{bmatrix}
    -1  & -1 & -1  \\
    -1  & 9  & -1  \\
    -1  & -1 & -1
\end{bmatrix}
\end{aligned}
\end{equation}
We perform convolution only to the positions marked as the extracted lines. In the next iteration, the sharpened image is fed back to the smoothing kernel with higher smoothing level, and we do the line extraction and sharpening convolution again until the maximal iteration is reached. The output from the iterative operation is a raw edge map of the last sharpened image extracted by the Canny edge extractor.

\begin{figure*}[!ht]
  \centerline{\includegraphics[width=13.65cm, height=3.9cm]{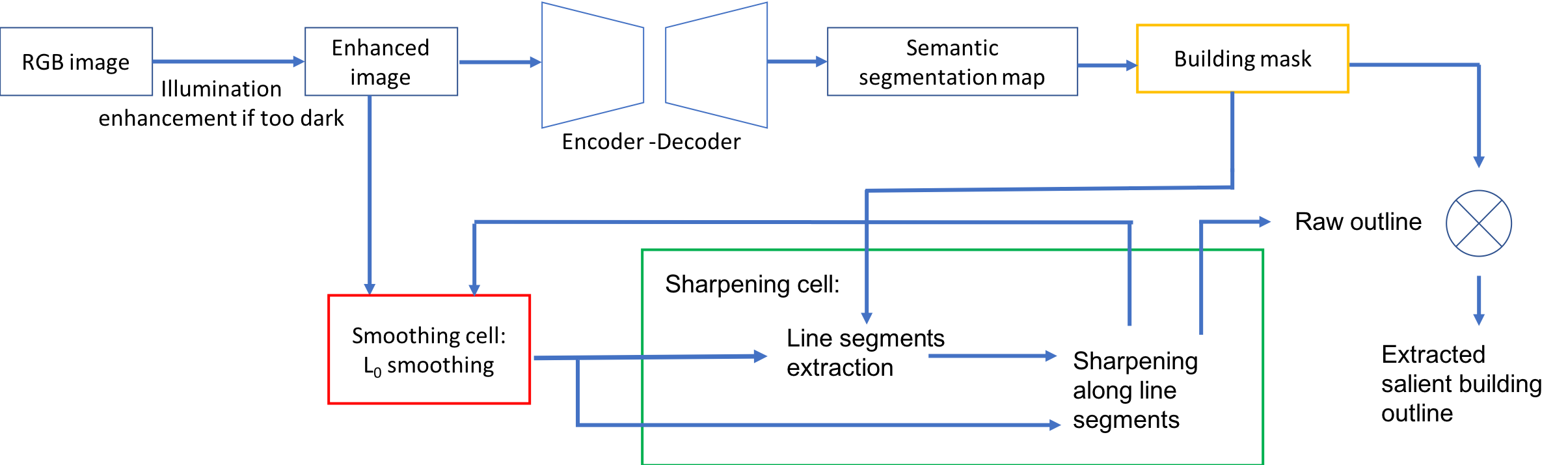}}
  \caption{The proposed Iterative Smoothing Line Enhancing (ISLE) framework.}
   \vspace{-0.3cm}
\end{figure*}

In order to obtain salient outlines only related to buildings, we create building masks from the semantic segmentation. We train an encoder-decoder structure network to do the semantic segmentation on the input image. The encoder network is a ResNet50 structure \cite{he2016deep}. The decoder network is a Pyramid Pooling Module of PSPNet \cite{zhao2017pyramid} with the dilated convolution \cite{ronneberger2015u}. We pretrain the network on the MIT ADE20K scene parsing dataset \cite{zhou2017scene}. This dataset contains 20K training images from general outdoor/indoor scenes with 150 different labels. We collect 6 kinds of labels related to buildings, "door", "building", "house", " wall", "awning", and "windowpane", to form the building masks. We also dilate the masks a little to gain some buffers to ensure all outlines are inside the mask. The examples of masks are in Fig. 2.

We apply the masks on the line extraction, to extract lines only within the predicted building area. Also, we apply the masks on the raw edge maps to obtain the refined salient building outline. The whole framework is in Fig. 3.

The iterative design is to prevent the strong oversmoothing to smooth out weak edges. Through iterations, from the low smoothing level to the high smoothing level, high frequency parts in images could be smoothed out and straight line structures could be retained simultaneously. Also weak edges that could not be detected by the Canny edge detector in the original images could also be enhanced. 

As illustrated in Fig. 1, observe from (b) and (e), one can see that the right side structures are intrinsically weak edges in the original images, and the left side structures are weak edges that could be smoothed out with a higher level smoothing. Also the crack is enlarged by the oversmoothing effect. However, using our proposed framework, one can observe from (h) that weak edges are enhanced and retained after iterative smoothing and line enhancing, and the crack is sealed.


\section{Experiments}
\label{sec:experiment}
\vspace{-0.5cm}
\begin{figure}[!htb]
  \centerline{\includegraphics[width=7.0cm]{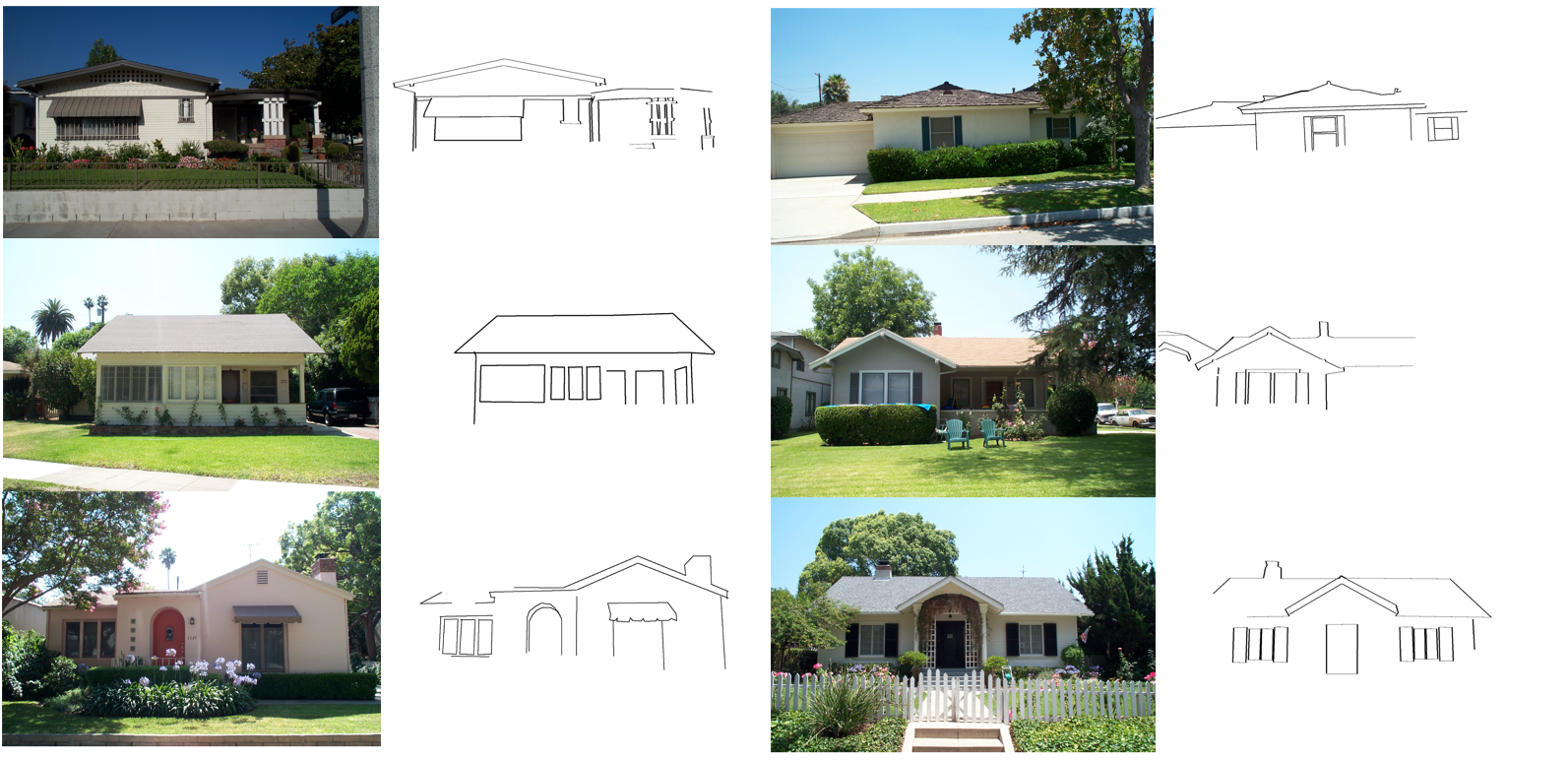}}
  \caption{Examples of the building images in Pasadena House dataset and drawn groundtruths.}
  \vspace{-0.3cm}
\end{figure}

To perform evaluation, we use Caltech Pasadena House dataset \cite{pasadena}. This dataset contains 240 house images. We randomly pick 50 house images and manually draw the salient outlines as the groundtruths. We draw the groundtruths based on following rules.

I. We draw outlines that are strong to human vision. Therefore, we do not draw outlines with low contrast. II. We ignore the occluded parts, since the edge information is lost there. III. We do not draw outlines outside buildings themselves, such as fences or outer walls. We show some examples of the drawn groundtruths in Fig. 4.

To evaluate the performance, we use Canny edge extractor with low and high threshold = (0.05,0.4), $L_0$ smoothing with higher smoothing level $\lambda = 0.03$, OEF \cite{hallman2015oriented}, Bayesian Salient Edge \cite{mukherjee2017salprop}, and RCF \cite{liu2019richer} as the comparison methods. OEF, Bayesian Salient Edge, and RCF are learning-based methods and use the large object salient outline extraction database BSD500 \cite{amfm_pami2011} for training. The output from the OEF, Bayesian Salient Edge, and RCF are probability edges in $[0,1]$. We binarize the edge with higher threshold, 0.72, 0.8, and 0.6 respectively, to attain higher level outline extraction. We tune all the mentioned parameters by the same person who draws the groundtruths. The criterion is to tune the parameter of the method when most of the extracted outlines are similar to the groundtruths, which follows the rules mentioned above.

Parameters of our method: $T_{high}=120$, $T_{low}=90$. High mode: Canny edge low and high threshold = (0.05,0.45) and maximal iteration is 7. Medium mode: Canny edge low and high threshold = (0.05,0.35) and maximal iteration is 5. We multiply pixel values with a factor 1.3. Low mode: Canny edge low and high threshold = (0.05,0.3) and maximal iteration is 3. We multiply pixel values with a factor 1.5. $\lambda$ sequence is $[0.001\; 0.003\; 0.005\; 0.008\; 0.01\; 0.02\; 0.03]$. The range of $\lambda$ is in $[0.001\; 0.1]$ as recommended in the original paper. Empirically we tune $\lambda$ sequence to have the best performance. We set the first 4 values on the scale of $10^{-3}$, low smoothing level. We set the later 3 values on the scale of $10^{-2}$, medium smoothing level.  

We show our results with visual comparisons in Fig. 5. We also give numerical comparisons using the structural similarity index (SSIM) between the results and the groundtruths in Table 1. To perform fair comparisons, we also apply the generated building masks on comparison methods and evaluate the results. From the figures, one can see that our visual results are most clean but could retain salient structures, and are most similar to the groundtruths overall. The numerical results also favor our proposed method. One may notice that OEF, RCF, and Bayesian Salient Edge are trained on the dataset for objects, not specifically for buildings. However, these are learning-based methods, and currently this new task lacks of large training and testing dataset. Thus, the disadvantage of learning-based methods appears when we work on the salient building outline extraction. We also release the 50 drawn groundtruths as the evaluation dataset \cite{dataset}.

We also conduct experiments on street building images collected from Google Map. We provide some visual results in Fig. 6 and Fig. 7, showing the ability of our method on another data source. One can observe that the proposed ISLE could extract salient outlines and exclude many details. More results could be found at \cite{dataset}.

\begin{table}[!tb]
\caption{Comparisons on the building outline extraction using the SSIM. RCF produces thick edges and thus scores low.}
\label{Table1}
\vspace{-0.5cm}
\begin{center}
\begin{tabular}{|c|c|c|c|}
\hline
         & Canny Edge& $L_0$ smoothing &  OEF         \\ \hline
SSIM     & 0.8839   & 0.8922  & 0.8907            \\ \hline
         & Bayesian Salient Edge & Proposed ISLE & RCF \\ \hline
SSIM     & 0.8949 & 0.8979 &  0.8685                 \\ \hline
\end{tabular}
\end{center}
\vspace{-0.5cm}
\end{table}

\begin{figure*}[!ht]
  \centerline{\includegraphics[width=17.4cm, height=8.7cm]{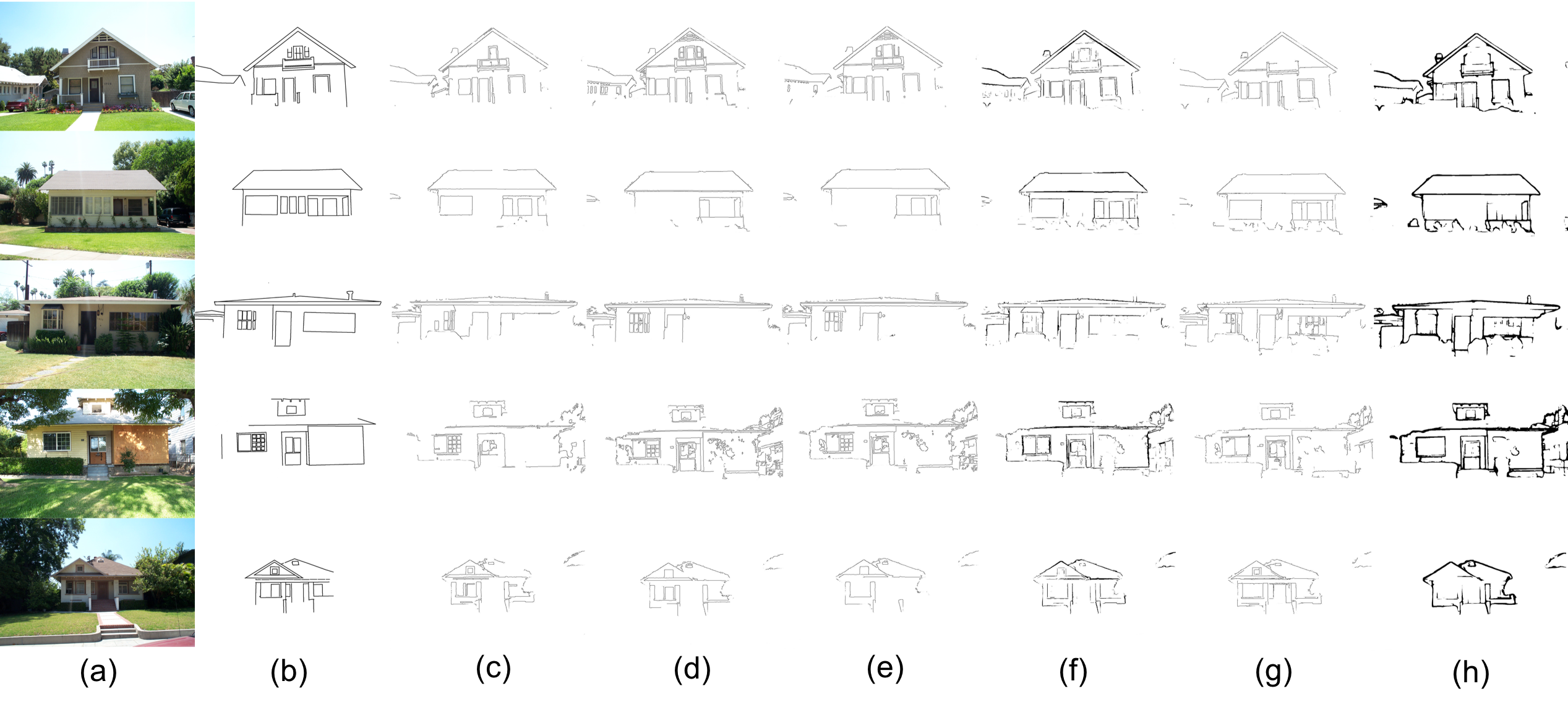}}
  \caption{(a) The example images in the Caltech Pasadena house dataset (b) The manually drawn groundtruths. (c) Outlines using the proposed ISLE framework. (d) Outlines using Canny edge extractor. (e) Outlines using $L_0$ norm smoothing. (f) Outlines using OEF. (g) Outlines using the Bayesian Salient Edge. (h) Outlines using RCF.}
\end{figure*}

\begin{figure}[!htb]
  \centerline{\includegraphics[width=5.84cm]{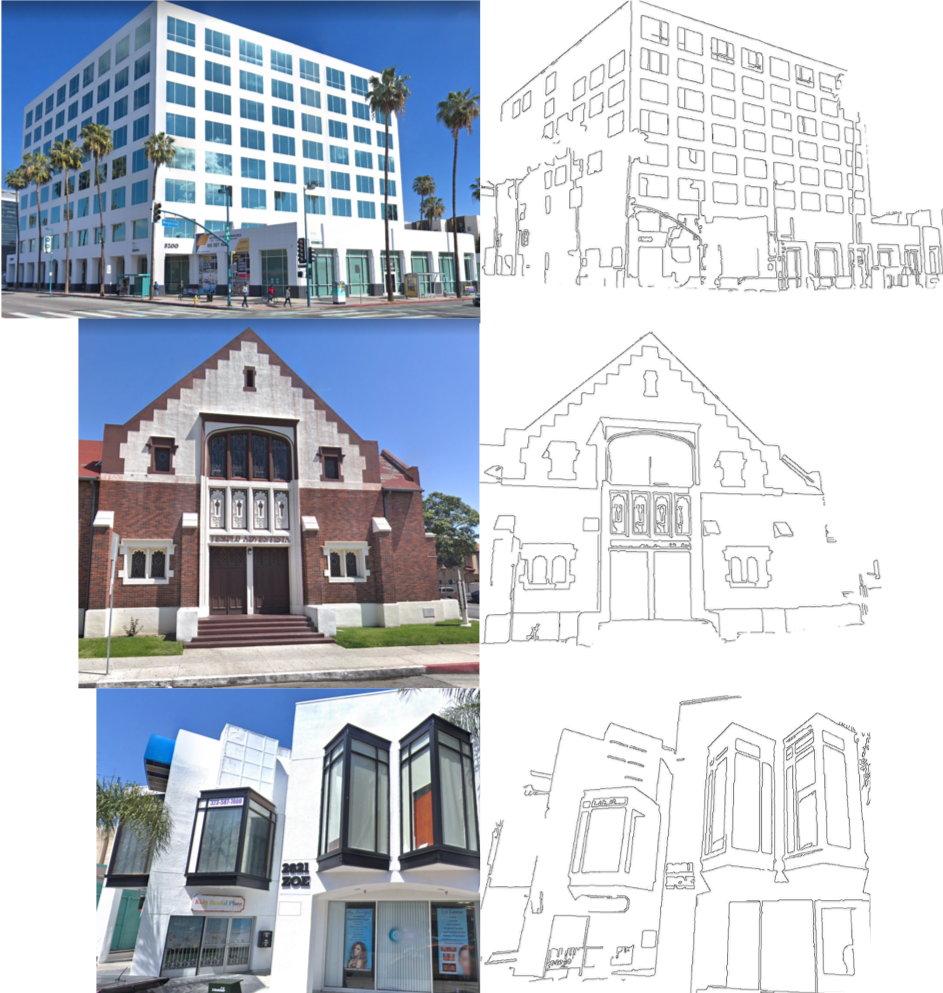}}
  \caption{Examples using images from Google Map with the proposed ISLE framework. (Part 1)}
  \vspace{-0.3cm}
\end{figure}

\begin{figure}[!htb]
  \centerline{\includegraphics[width=5.50cm]{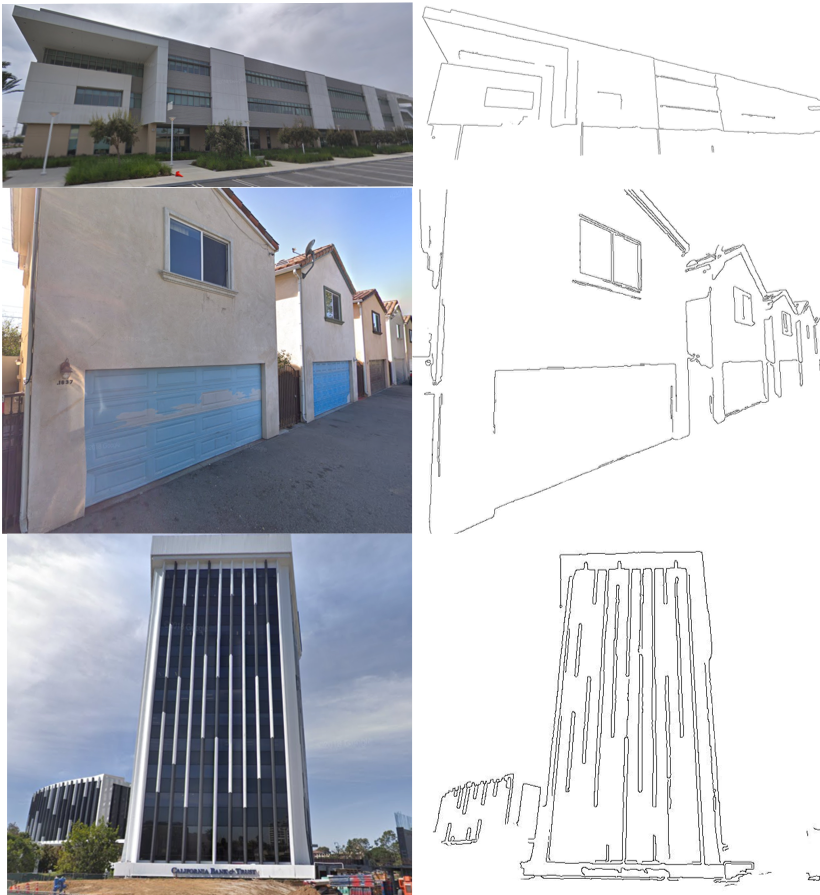}}
  \caption{Examples using images from Google Map with the proposed ISLE framework. (Part 2)}
  \vspace{-0.3cm}
\end{figure}



\section{Conclusion}
\label{sec:con}
In this paper, we address building outline enhancement and extraction from images taken from consumer cameras. We propose an iterative $L_0$ smoothing and line enhancing framework to simultaneously retain line structures of buildings and smooth out high frequency parts of images. We also propose to use a building mask obtained from an encoder-decoder network. Experiments show the ability of our method. We also release the drawn evaluation dataset for the public to use.

\bibliographystyle{IEEEbib}
\bibliography{strings,refs}

\end{document}